\documentclass[aps,preprintnumbers,superscriptaddress,showpacs]{revtex4}
\usepackage{epsfig}
\usepackage{psfrag}
\usepackage{amsfonts}
\usepackage{graphicx}
\usepackage{dcolumn}
\usepackage{bm}

\begin{document}

\title{Effect of gluon condensate on holographic Schwinger effect}

\author{Zi-qiang Zhang}
\email{zhangzq@cug.edu.cn} \affiliation{School of Mathematics and
Physics, China University of Geosciences, Wuhan 430074, China}

\author{Xiangrong Zhu}
\email{xrongzhu@zjhu.edu.cn} \affiliation{School of Science,
Huzhou University, Huzhou 313000, China}

\author{De-fu Hou}
\email{houdf@mail.ccnu.edu.cn} \affiliation{Central China Normal
University, Wuhan 430079, China}

\begin{abstract}
We perform the potential analysis in holographic Schwinger effect
in a deformed anti-de Sitter (AdS) background with backreaction
due to the gluon condensate. We determine the potential by
analyzing the classical string action attaching on a probe
D3-brane sitting at an intermediate position in the bulk AdS
space. It is found that the inclusion of the gluon condensate
reduces the production rate, reverse to the effect of the
temperature. Also, we evaluate the critical electric field by
Dirac-Born-Infeld (DBI) action.

\end{abstract}

\maketitle
\section{Introduction}
It is generally accepted that vacuum in quantum field theory (QFT)
is not actually barren. Rather, it contains lots of virtual
particles and antiparticles due to quantum fluctuations. For
instance, in the vacuum of quantum electrodynamics (QED), virtual
electron-position pairs are supposed to be momentarily created and
annihilated. Moreover, these virtual particles could be
materialized and become real particles in a strong electric-field.
This non-perturbative phenomenon is known as the Schwinger effect.
The production rate $\Gamma$ (per unit time and volume) has been
evaluated by Schwinger for the case of weak-coupling and
weak-field in 1951 \cite{JS}
\begin{equation}
\Gamma\sim exp\Big({\frac{-\pi m^2}{eE}}\Big),
\end{equation}
where $E$, $m$ and $e$ are an external electric-field, an electron
mass and an elementary electric charge, respectively. Thirty one
years later, Affleck-Alvarez-Manton (AAM) generalized it to the
case of arbitrary-coupling and weak-field \cite{IK}
\begin{equation}
\Gamma\sim exp\Big({\frac{-\pi m^2}{eE}+\frac{e^2}{4}}\Big),
\end{equation}
From the above formulas of $\Gamma$, one finds there is no
critical field in the Schwinger case. While in the AAM case, there
is a critical field at $eE_c=(4\pi/e^2)m^2\simeq137m^2$, but it is
far beyond the weak-field condition $eE\ll m^2$. Thus, it seems
that one could not get the critical field under the weak-field
condition.

Actually, the Schwinger effect is not confined to QED but
ubiquitous for QFT coupled to an U(1) gauge field. However, it
remains difficult to tackle this issue with the standard method in
QFT. A possible way is to use the AdS/CFT correspondence
\cite{Maldacena:1997re,Gubser:1998bc,MadalcenaReview} by realizing
QFT (or rather confining gauge theories) with appropriate D-brane
set-up. In 2011, Semenoff and Zarembo proposed \cite{GW} that the
Schwinger effect could be modeled in the higgsed $\mathcal N=4$
supersymmetric Yang-Mills (SYM) theory. Specifically, a $\mathcal
N=4$ SYM theory system coupled with an U(1) gauge field can be
realized by breaking the gauge group from $SU(N+1)$ to
$SU(N)\times U(1)$ via the Higgs mechanism. In this approach, the
production rate and the critical electric field (at large $N$ and
large 't Hooft coupling $\lambda$) are evaluated as
\begin{equation}
\Gamma\sim
exp\Big[-\frac{\sqrt{\lambda}}{2}\Big(\sqrt{\frac{E_c}{E}}-\sqrt{\frac{E}{E_c}}\Big)^2\Big],\qquad
E_c=\frac{2\pi m^2}{\sqrt{\lambda}}, \label{gama}
\end{equation}
interestingly, the resulting critical field is completely
consistent with the DBI result. Subsequently, there are many
attempts to address the holographic Schwinger in this direction.
For instance, the potential analysis in holographic Schwinger
effect has been investigated in various backgrounds
\cite{YS,YS1,YS2,KB,MG,ZQ,LS}. The holographic Schwinger effect
and negative differential conductivity have been discussed in
\cite{SCH}. The holographic Schwinger effect with constant
electric and magnetic fields was considered in \cite{SB,YS3}. For
a study of this quantity in de Sitter spacetime, see \cite{WF}.
Moreover, the holographic Schwinger effect has been analyzed from
the imaginary part of a probe brane action \cite{KHA,KHA1,XW,KG}.
For a recent review on this topic, see \cite{DK}.

The aim of this paper is to study the effect of the gluon
condensate on the (holographic) Schwinger effect. The gluon
condensate was proposed in \cite{mas} as a measure for
nonperturbative physics in QCD (at zero temperature).
Subsequently, it was regarded as an order parameter for
(de)confinement and used to explore the nonperturbative natures of
quark gluon plasma (QGP) \cite{gl0,gl1,gl2,g13}. Moreover, lattice
results show that the gluon condensate is non-zero at high
temperature, in particular, its value drastically changes near
$T_c$ (the critical temperature of the deconfinement transition)
regardless of the number of quark flavors \cite{gb}. Due to the
above reasons, it would be natural and very interesting to study
the possible effect that the gluon condensate might cause on
various observables or quantities. Recently, there has been such
research from holography. For instance, the effect of the gluon
condensate on the heavy quark potential was studied in \cite{yk}
and it was shown that the potential becomes deeper as the value of
the gluon condensate decreases. Also, the gluon condensate
dependence of the jet quenching parameter and drag force was
considered in \cite{zq} and it was found that the inclusion of the
gluon condensate increases the energy loss. Not long ago, the
authors of \cite{YQ} analyzed the effect of the gluon condensate
on the imaginary potential and found the dropping gluon condensate
reduces the absolute value of imaginary potential thus decreasing
the thermal width. Motivated by this, in this paper we study the
effect of the gluon condensate on the Schwinger effect. More
immediately, we want to understand how the gluon condensate
affects the production rate. Also, this work could be considered
as the generalization of \cite{YS} to the case with gluon
condensation.

The organization of the paper is as follows. In the next section,
we introduce the deformed AdS backgrounds with backreaction due to
the gluon condensate. In section 3, we perform the potential
analysis for the Schwinger effect in these backgrounds and discuss
how the gluon condensate modifies the production rate. The
conclusions and discussions are given in section 4.

\section{Setup}
The 5-dimensional (5D) gravity action (in Minkowski) with a
dilaton coupled is given by \cite{sn}
\begin{equation}
I=\frac{1}{2\kappa^2}\int
d^5x\sqrt{g}(\mathcal{R}+\frac{12}{R^2}-\frac{1}{2}\partial_M\phi\partial^M\phi),\label{action}
\end{equation}
where $\kappa^2$ is the 5D Newtonian constant. $\mathcal{R}$
denotes the Ricci scalar. $R$ represents the AdS curvature
(hereafter we set $R=1$). $\phi$ refers to the dilaton, coupled to
the gluon operator. By solving the Einstein equation and the
dilaton equation of motion, one can obtain two relevant solutions.
The first is the dilaton-wall solution, given by \cite{ak,cc}
\begin{equation}
ds^2=r^2\sqrt{1-c^2r^{-8}}(d\vec{x}^2-dt^2)+\frac{dr^2}{r^2},\label{so1}
\end{equation}
and the corresponding dilaton profile is
\begin{equation}
\phi(r)=\sqrt{\frac{3}{2}}\log
(\frac{1+cr^{-4}}{1-cr^{-4}})+\phi_0,
\end{equation}
where $\vec{x}=x_1,x_2,x_3$ are the boundary coordinates. $r$
describes the 5D coordinate and the boundary is $r=\infty$.
$\phi_0$ denotes a constant. $c$ represents the gluon
condensation.

Another is the dilaton black hole solution, given by \cite{db,yk1}
\begin{equation}
ds^2=r^2H(r)d\vec{x}^2-r^2P(r)dt^2+\frac{dr^2}{r^2},\label{so2}
\end{equation}
with
\begin{eqnarray}
H(r)&=&(1+fr^{-4})^{(f+a)/{2f}}(1-fr^{-4})^{(f-a)/{2f}},\nonumber\\
P(r)&=&(1+fr^{-4})^{(f-3a)/{2f}}(1-fr^{-4})^{(f+3a)/{2f}},\nonumber\\
f^2&=&a^2+c^2,\nonumber\\
\end{eqnarray}
and
\begin{equation}
\phi(r)=\frac{c}{f}\sqrt{\frac{3}{2}}\log
(\frac{1+fr^{-4}}{1-fr^{-4}})+\phi_0.
\end{equation}

As discussed in \cite{yk1}, the solution (\ref{so2}) is well
defined only in the range $r_f<r<\infty$ with $r_f\equiv f^{1/4}$,
where $r_f$ could be considered as the IR cut-off. The parameter
$f$ determines the position of the singularity. $a$ is related to
the temperature as $a=(\pi T)^4/4$. Note that for $a=0$,
(\ref{so2}) reduces to the dilaton-wall solution, and for $c=0$ it
becomes the Schwarzschild black hole solution. Moreover, there is
a Hawking-Page transition between (\ref{so1}) and (\ref{so2}) at
some critical value of $a$. Therefore, the dilaton-wall solution
is for the confined phase and the dilaton black hole solution is
for the deconfined phase. For more details about the two
solutions, we refer to \cite{yk1}.

\section{potential analysis in Schwinger effect}
In this section we follow the approach in \cite{YS} to study the
effect of the gluon condensate on the Schwinger effect. Since the
dilaton-wall background could be derived from the dilaton black
hole background by plugging $a=0$ in (\ref{so2}), we will perform
(only) the potential analysis for the latter but discuss the
results for both.

\subsection{Coulomb potential and static energy}
One considers a rectangular Wilson loop on the probe D3-brane
located at $r=r_0$ and impose the following ansatz
\begin{equation}
t=\tau, \qquad x_1=\sigma, \qquad x_2=0,\qquad x_3=0, \qquad
r=r(\sigma). \label{par}
\end{equation}
The Nambu-Goto action is
\begin{equation}
S=T_F\int d\tau d\sigma\mathcal L=T_F\int d\tau d\sigma\sqrt{g},
\qquad T_F=\frac{1}{2\pi\alpha^\prime} \label{S}
\end{equation}
where $T_F$ denotes the string tension. $\alpha^\prime$ is related
to $\lambda$ by
$\frac{R^2}{\alpha^\prime}=\frac{1}{\alpha^\prime}=\sqrt{\lambda}$.
$g$ denotes the determinant of the induced metric with
\begin{equation}
g_{\alpha\beta}=g_{\mu\nu}\frac{\partial
X^\mu}{\partial\sigma^\alpha} \frac{\partial
X^\nu}{\partial\sigma^\beta},
\end{equation}
where $g_{\mu\nu}$ and $X^\mu$ are the metric and target space
coordinates, respectively.

Plugging (\ref{par}) into (\ref{so2}), the Lagrangian reads
\begin{equation}
\mathcal L=\sqrt{A(r)+B(r)(\frac{dr}{d\sigma})^2},\label{L}
\end{equation}
with
\begin{equation}
A(r)=r^4H(r)P(r)e^{\phi(r)},\qquad B(r)=P(r)e^{\phi(r)}.
\end{equation}

Since $\mathcal L$ does not depend on $\sigma$ explicitly, the
corresponding Hamiltonian is a constant
\begin{equation}
\mathcal{H}=\mathcal L-\frac{\partial\mathcal
L}{\partial(\frac{dr}{d\sigma})}(\frac{dr}{d\sigma})=Constant.
\end{equation}

Imposing the boundary condition at $\sigma=0$,
\begin{equation}
\frac{dr}{d\sigma}=0,\qquad  r=r_c\qquad (r_f<r_c<r_0)\label{con},
\end{equation}
given that, one has
\begin{equation}
\frac{dr}{d\sigma}=\sqrt{\frac{A^2(r)-A(r)A(r_c)}{A(r_c)B(r)}}\label{dotr},
\end{equation}
where $A(r_c)=A(r)|_{r=r_c}$.

Integrating (\ref{dotr}), the inter-distance between the
$q\bar{q}$ (test particles) can be written as
\begin{equation}
x=2\int_{r_c}^{r_0}dr\sqrt{\frac{A(r_c)B(r)}{A^2(r)-A(r)A(r_c)}}\label{xx}.
\end{equation}

On the other hand, plugging (\ref{L}) and (\ref{dotr}) into
(\ref{S}), the sum of Coulomb potential and static energy of the
$q\bar{q}$ is expressed as
\begin{equation}
V_{CP+E}=2T_F\int_{r_c}^{r_0}dr\sqrt{\frac{A(r)B(r)}{A(r)-A(r_c)}}.\label{en}
\end{equation}

\subsection{critical electric field}
Next, we calculate the critical field. The DBI action takes the
form
\begin{equation}
S_{DBI}=-T_{D3}\int
d^4x\sqrt{-det(G_{\mu\nu}+\mathcal{F}_{\mu\nu})}\label{dbi},
\end{equation}
where
\begin{equation}
T_{D3}=\frac{1}{g_s(2\pi)^3\alpha^{\prime^2}}, \qquad
\mathcal{F}_{\mu\nu}=2\pi\alpha^\prime F_{\mu\nu},
\end{equation}
with $T_{D3}$ the D3-brane tension.

Applying  (\ref{so2}) and assuming the electric field is turned on
along the $x_1$-direction \cite{YS}, one has
\begin{equation}
G_{\mu\nu}+\mathcal{F}_{\mu\nu}=\left(
\begin{array}{cccc}
-r^2P(r)e^{\phi(r)/2} & 2\pi\alpha^\prime E & 0 & 0\\
 -2\pi\alpha^\prime E & r^2H(r)e^{\phi(r)/2} & 0 & 0 \\
 0 & 0 & r^2H(r)e^{\phi(r)/2} & 0\\
0 & 0 & 0 & r^2H(r)e^{\phi(r)/2}
\end{array}
\right),
\end{equation}
which gives
\begin{equation}
det(G_{\mu\nu}+\mathcal{F}_{\mu\nu})=r^4H^2(r)e^{\phi(r)}[(2\pi\alpha^\prime)^2E^2-r^4P(r)H(r)e^{\phi(r)}].\label{det}
\end{equation}

Putting (\ref{det}) into (\ref{dbi}) and making the D3-brane
located at $r=r_0$, one obtains
\begin{equation}
S_{DBI}=-T_{D3}r_0^2H(r_0)e^{\phi(r_0)/2}\int d^4x
\sqrt{r_0^4P(r_0)H(r_0)e^{\phi(r_0)}-(2\pi\alpha^\prime)^2E^2}\label{dbi1},
\end{equation}
where $P(r_0)=P(r)|_{r=r_0}$, etc.

The quantity under the square root of (\ref{dbi1}) should be
non-negative, yielding
\begin{equation}
r_0^4P(r_0)H(r_0)e^{\phi(r_0)}-(2\pi\alpha^\prime)^2E^2\geq0,\label{ec}
\end{equation}
resulting in
\begin{equation}
E\leq\frac{1}{2\pi\alpha^\prime}r_0^2\sqrt{P(r_0)H(r_0)e^{\phi(r_0)}}\equiv
T_Fr_0^2\sqrt{P(r_0)H(r_0)e^{\phi(r_0)}}.
\end{equation}

At last, one arrives at the critical field
\begin{equation}
E_c=T_Fr_0^2\sqrt{P(r_0)H(r_0)e^{\phi(r_0)}},\label{ec1}
\end{equation}
one can see that $E_c$ depends on the temperature as well as the
gluon condensate.

\subsection{total potential}
The remaining task is to compute the total potential, which takes
the form
\begin{eqnarray}
V_{tot}(x)&=&V_{CP+E}-Ex\nonumber\\&=&2T_F\int_{r_c}^{r_0}dr\sqrt{\frac{A(r)B(r)}{A(r)-A(r_c)}}\nonumber\\&-&
2\alpha T_Fr_0^2\sqrt{P(r_0)H(r_0)e^{\phi(r_0)}}
\int_{r_c}^{r_0}dr\sqrt{\frac{A(r_c)B(r)}{A^2(r)-A(r)A(r_c)}}.
\label{V}
\end{eqnarray}
where $\alpha\equiv\frac{E}{E_c}$. It seems quite difficult to
evaluate the above expression analytically, but it is possible
numerically. To ensure stable numerics, it turns out to be more
convenient to use the following dimensionless parameters like
\begin{equation}
y\equiv\frac{r}{r_c},\qquad m\equiv\frac{r_c}{r_0},\qquad
\end{equation}
given that, (\ref{V}) becomes
\begin{eqnarray}
V_{tot}(x)&=&V_{CP+E}-Ex\nonumber\\&=&2T_Fmr_0\int_{1}^{1/m}dy\sqrt{\frac{A(y)B(y)}{A(y)-A(y_c)}}\nonumber\\&-&
2\alpha T_Fmr_0^3\sqrt{P(r_0)H(r_0)e^{\phi(r_0)}}
\int_{1}^{1/m}dy\sqrt{\frac{A(y_c)B(y)}{A^2(y)-A(y)A(y_c)}}.
\label{V1}
\end{eqnarray}
where
\begin{eqnarray}
A(y)&=&(mr_0y)^4H(y)P(y)e^{\phi(y)},\qquad A(y_c)=(mr_0)^4H(y_0)P(y_0)e^{\phi(y_0)},\nonumber\\
B(y)&=&P(y)e^{\phi(y)},\qquad
\phi(y)=\frac{c}{f}\sqrt{\frac{3}{2}}\log(\frac{1+f(mr_0y)^{-4}}{1-f(mr_0y)^{-4}})+\phi_0,\nonumber\\
H(y)&=&(1+f(mr_0y)^{-4})^{\frac{f+a}{2f}}(1-f(mr_0y)^{-4})^{\frac{f-a}{2f}},\nonumber\\
P(y)&=&(1+f(mr_0y)^{-4})^{\frac{f-3a}{2f}}(1-f(mr_0y)^{-4})^{\frac{f+3a}{2f}},
\end{eqnarray}
with $H(y_0)\equiv H(r)|_{r=mr_0}$, $P(y_0)\equiv P(r)|_{r=mr_0}$
and $\phi(mr_0)\equiv \phi(r)|_{r=mr_0}$. One can check that by
turning off the gluon condensate effect in (\ref{V1}), the results
of SYM case \cite{YS} are recovered (note that the temperature
formula in this paper without gluon condensate is $r_f=\pi
T/\sqrt{2}$ but that in \cite{YS} is $r_h=\pi T$).

\subsection{dilaton-wall result}
\begin{figure}
\centering
\includegraphics[width=8cm]{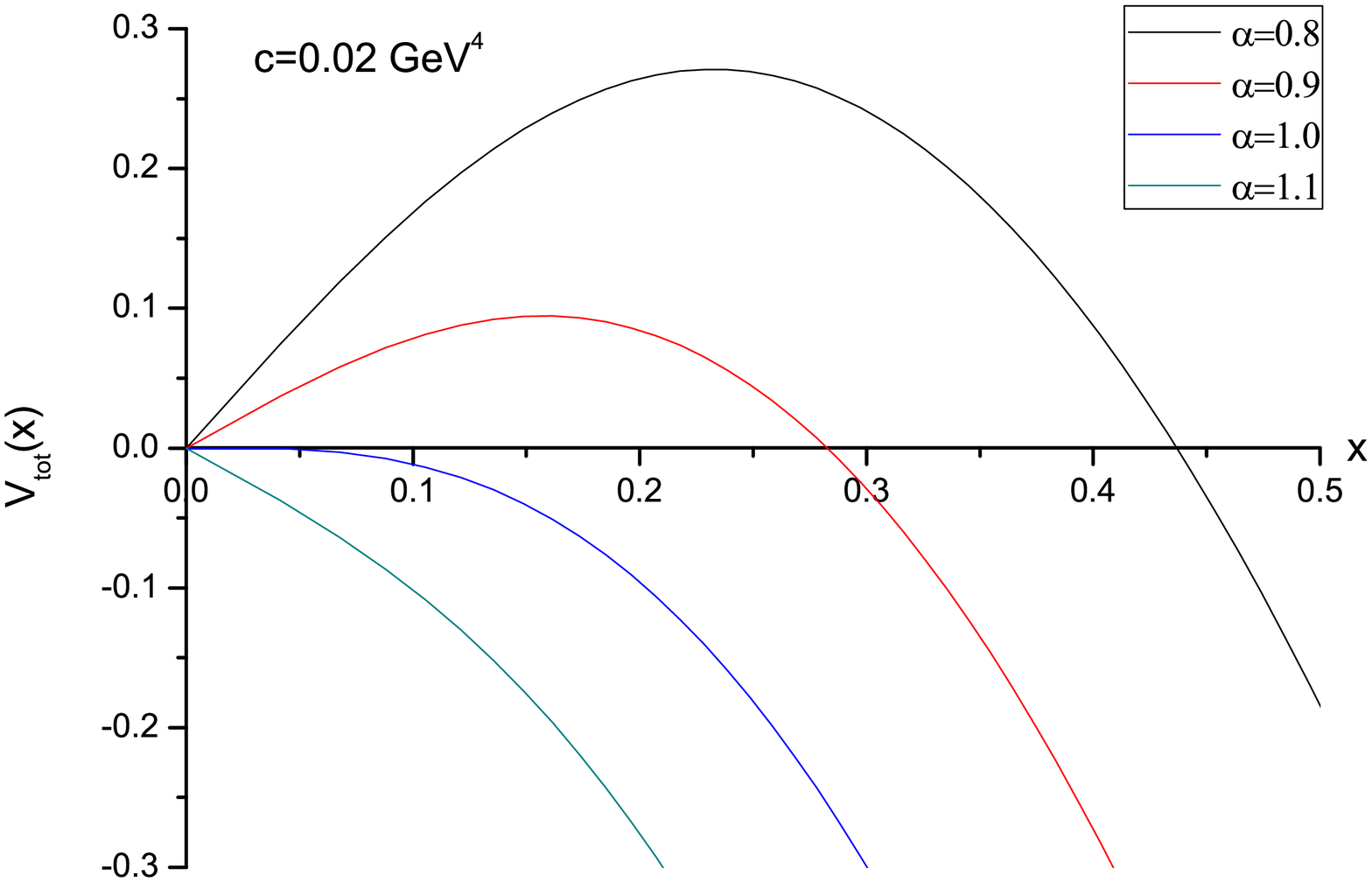}
\includegraphics[width=8cm]{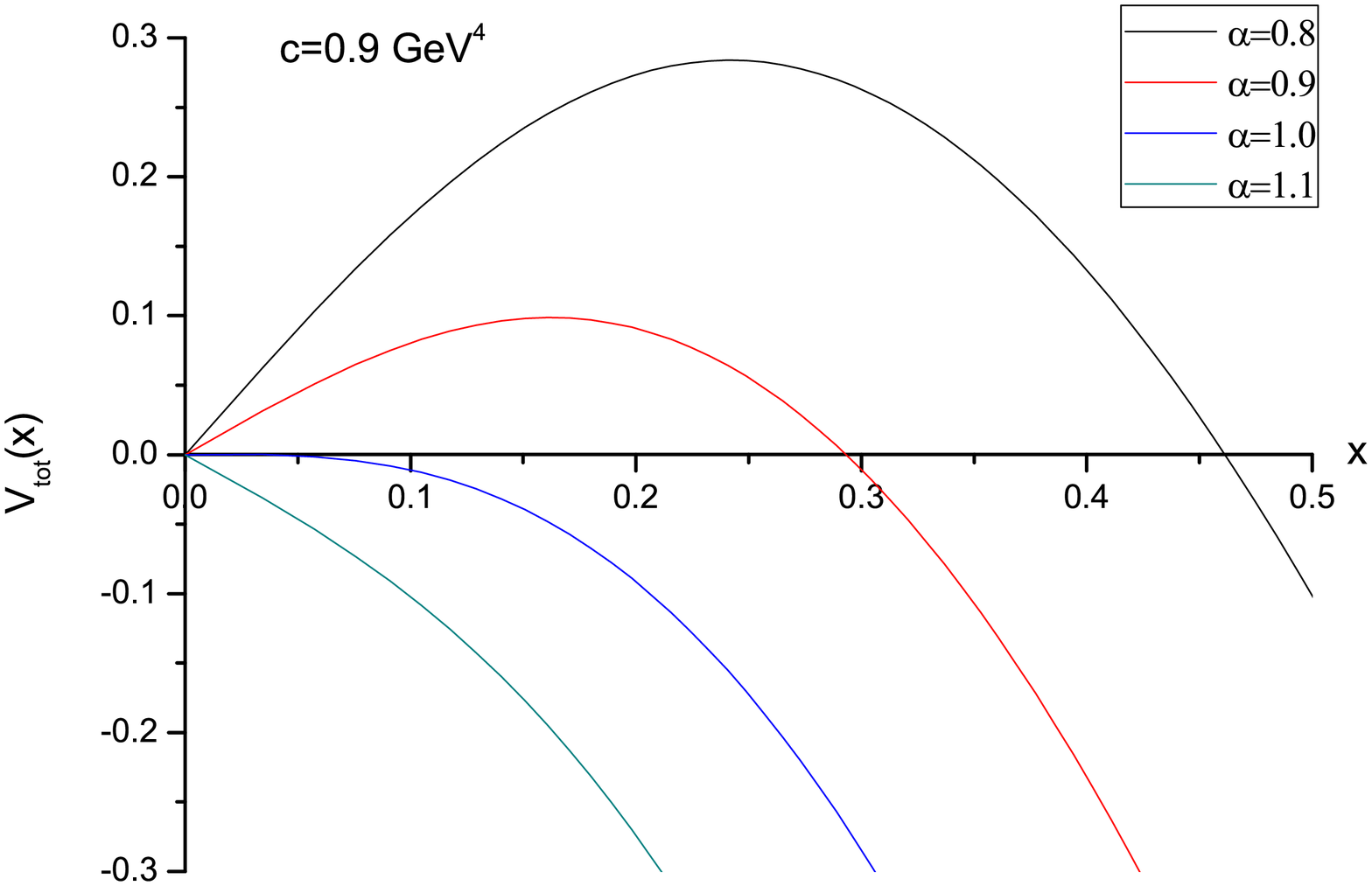}
\caption{$V_{tot}(x)$ versus $x$ with different $c$ for the
dilaton-wall background. In both panels from top to bottom
$\alpha=0.8, 0.9, 1.0, 1.1$, respectively.}
\end{figure}

\begin{figure}
\centering
\includegraphics[width=8cm]{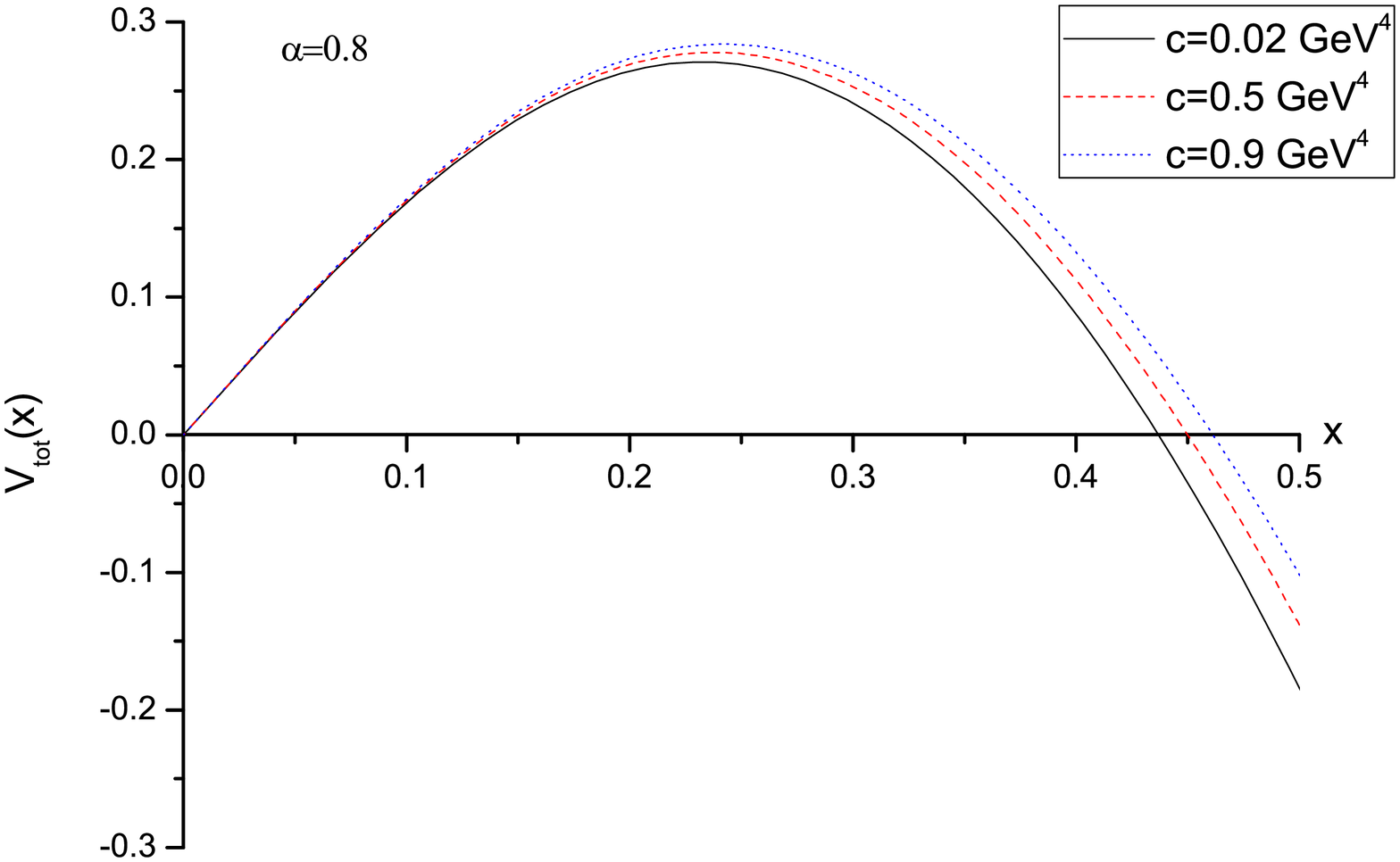}
\includegraphics[width=8cm]{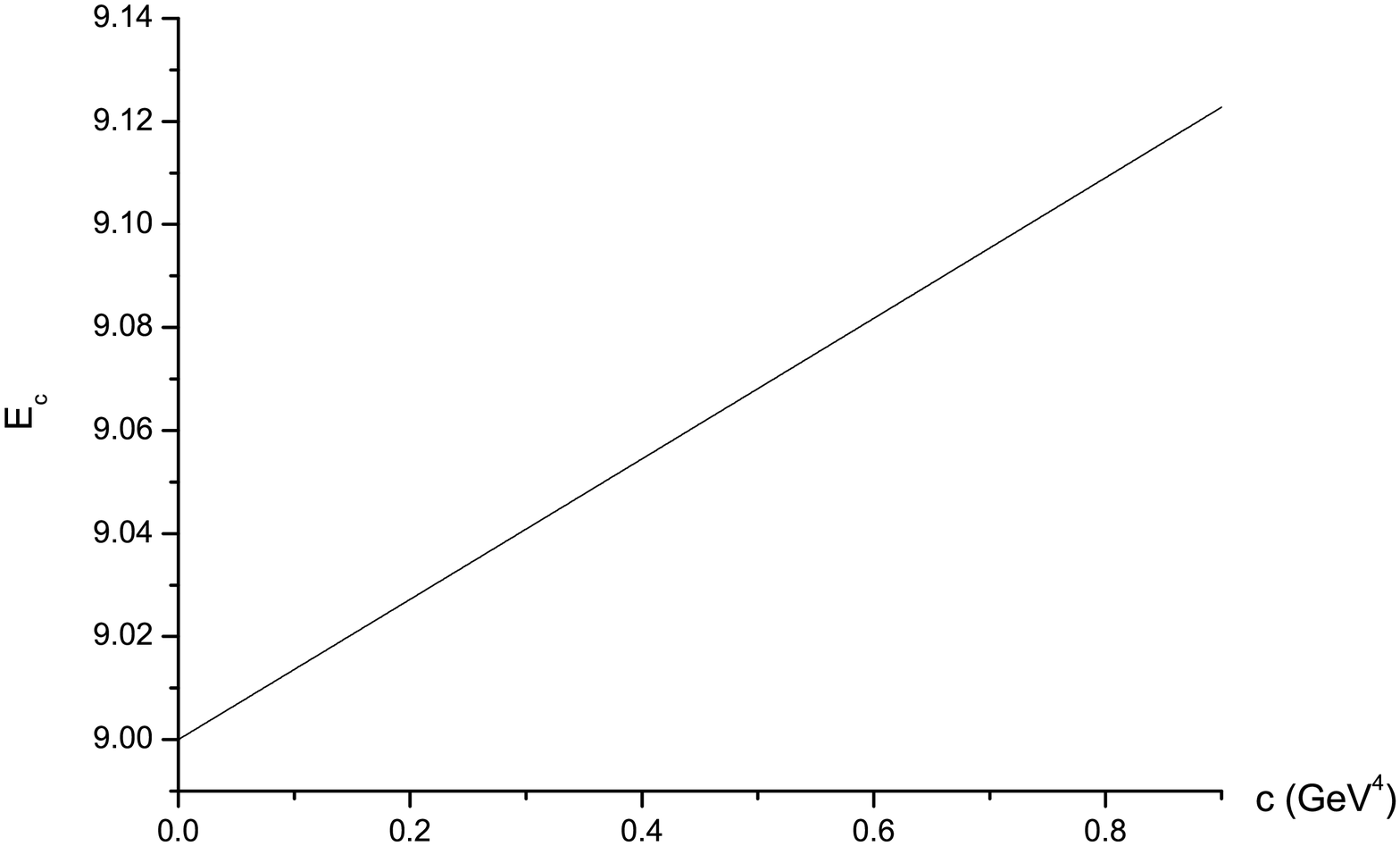}
\caption{Left: $V_{tot}(x)$ versus $x$ with fixed $\alpha=0.8$ and
different $c$ for the dilaton-wall background. From top to bottom
$c=0.9, 0.5, 0.02$ $GeV^4$, respectively. Right: $E_c$ versus
$c$.}
\end{figure}

Before numerical computation, we determine the values of some
parameters. First, we set $T_F=1$ and choose an appropriate value
of $r_0$, e.g., $r_0=3$, similar to \cite{YS}. Also, we take
$0\leq c\leq0.9 GeV^4$ and $\phi_0=0$, as follows from
\cite{yk,zq}.

We first discuss the results for the dilaton-wall background (zero
temperature case). In fig.1, we plot $V_{tot}(x)$ against $x$ with
different values of $c$, where the left panel is for $c=0.02
GeV^4$ (small gluon condensate) while the right one $c=0.9 GeV^4$
(large gluon condensate). Other cases with different values of $c$
have similar picture. From these figures, one can see that there
are mainly three situations: When $E<E_c$ ($\alpha<1$), the
potential barrier is present and the Schwinger effect can occur as
a tunneling process. As $E$ increases, the potential barrier
decreases gradually and vanishes at $E=E_c$ ($\alpha=1$). When
$E>E_c$ ($\alpha>1$), the system becomes catastrophically
unstable. The above analysis are in agreement with \cite{YS}.

In order to study how the gluon condensate influences the
Schwinger effect, we plot $V_{tot}(x)$ versus $x$ with fixed
$\alpha=0.8$ for different values of $c$ in the left panel of
fig.2. One can see that as $c$ increases, the height and width of
the potential barrier both increase. As you know, the higher (or
the wider) the potential barrier, the harder the produced pair
escapes to infinity. Therefore, one concludes that the presence of
the gluon condensate increases the potential barrier thus
decreasing the Schwinger effect.

Also, one can analyze the effect of the gluon condensate on the
critical field. To this end, we plot $E_c$ versus $c$ in the right
panel of fig.2. One finds that increases $c$ leads to increasing
$E_c$ thus making the Schwinger effect harder, consistently the
previous potential analysis.

\subsection{dilaton black hole result}
Next, we discuss the results for the dilaton black hole background
(finite temperature case). Likewise, the findings are presented in
form of plots, i.e, fig.3$\sim$fig.5, where fig.3 shows the
general behavior of the potential for various $T$ and fixed $c$
(other cases with different values of $c$ have similar picture).
One can see that there are still three cases for the potential,
similar to the dilaton-wall case.

In order to see how the gluon condensate modifies the Schwinger
effect at non-zero temperature, we plot $V_{tot}(x)$ versus $x$
with fixed $T$ and different values of $c$ in the left panel of
fig.4. One gets similar results: the inclusion of the gluon
condensate increases the potential barrier thus decreasing the
Schwinger effect. Also, the same conclusion could be obtained from
the gluon condensate dependence of $E_c$ (see the right panel of
fig.4): $E_c$ increases with $c$. Interestingly, it was argued
\cite{LS} that the D-instanton density (corresponds to the vacuum
expectation value of the gluon condensation) decreases the
Schwinger effect as well.

Furthermore, to understand the temperature dependence of the
Schwinger effect, we plot $V_{tot}(x)$ versus $x$ with different
$T$ (as well as $E_c$ versus $T$) in fig.5. From the left panel,
one can see that at fixed $c$, increasing $T$ leads to decreasing
the potential barrier, while from the right panel one finds $E_c$
decreases with $T$, which means increasing $T$ enhances the
Schwinger effect. Therefore, the gluon condensate and temperature
have opposite effects on the Schwinger effect. The physical
significance of the results will be discussed in the next section

\begin{figure}
\centering
\includegraphics[width=8cm]{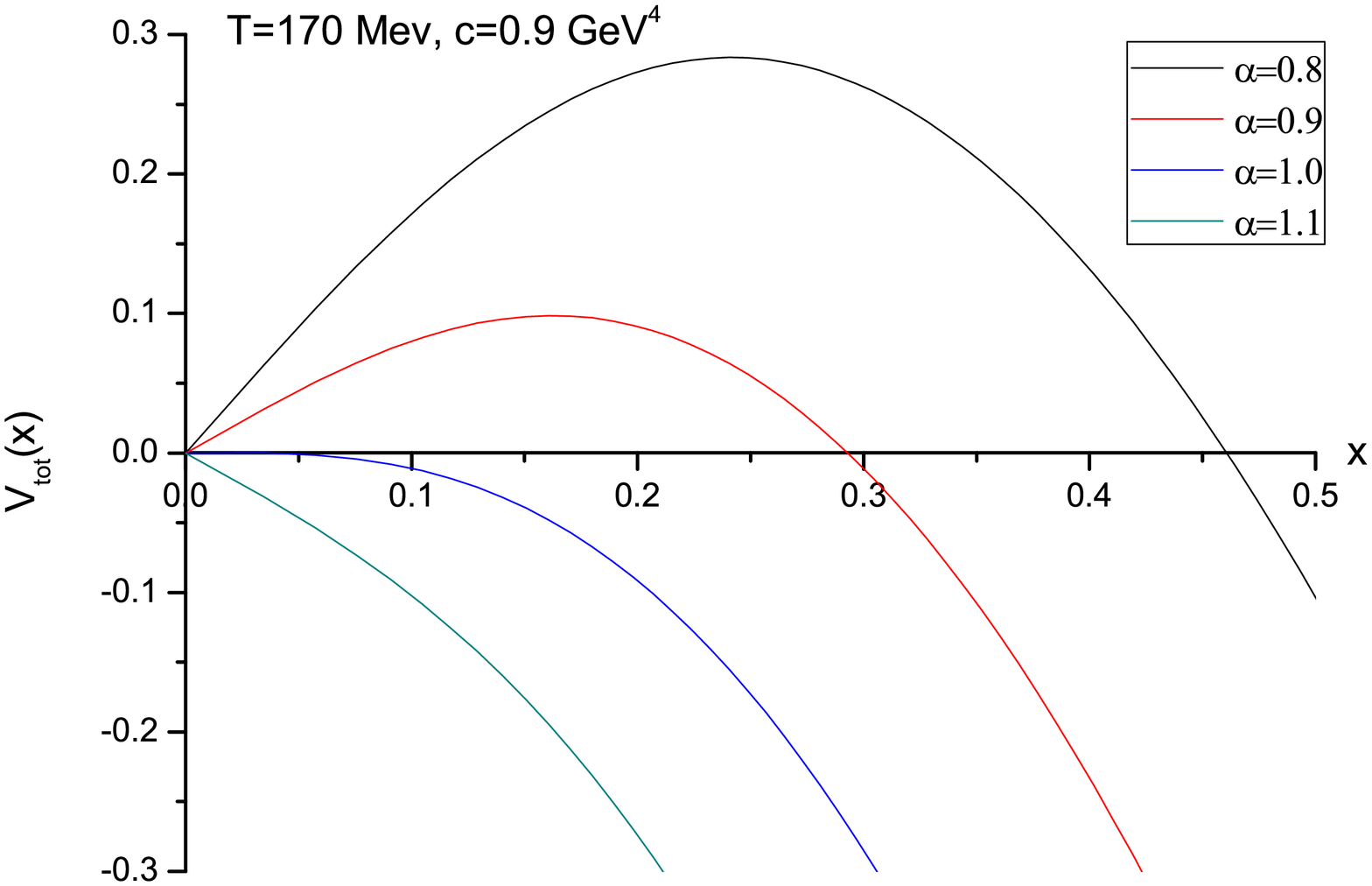}
\includegraphics[width=8cm]{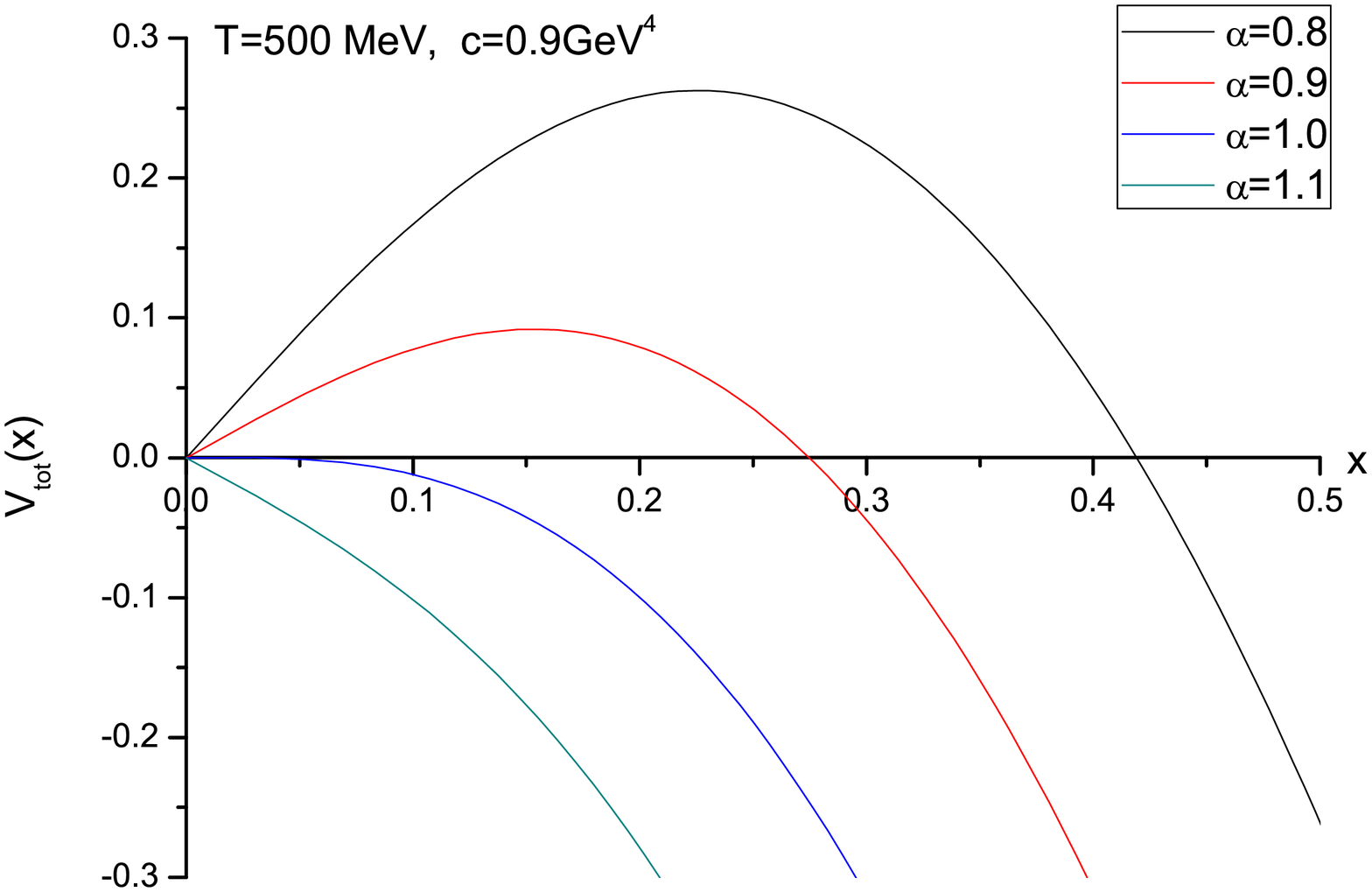}
\caption{$V_{tot}(x)$ versus $x$ with fixed $c=0.9 GeV^4$ and
different $T$ for the dilaton black hole background. Left: $T=170
MeV$; Right: $T=500 MeV$. In both panels from top to bottom
$\alpha=0.8, 0.9, 1.0, 1.1$, respectively.}
\end{figure}

\begin{figure}
\centering
\includegraphics[width=8cm]{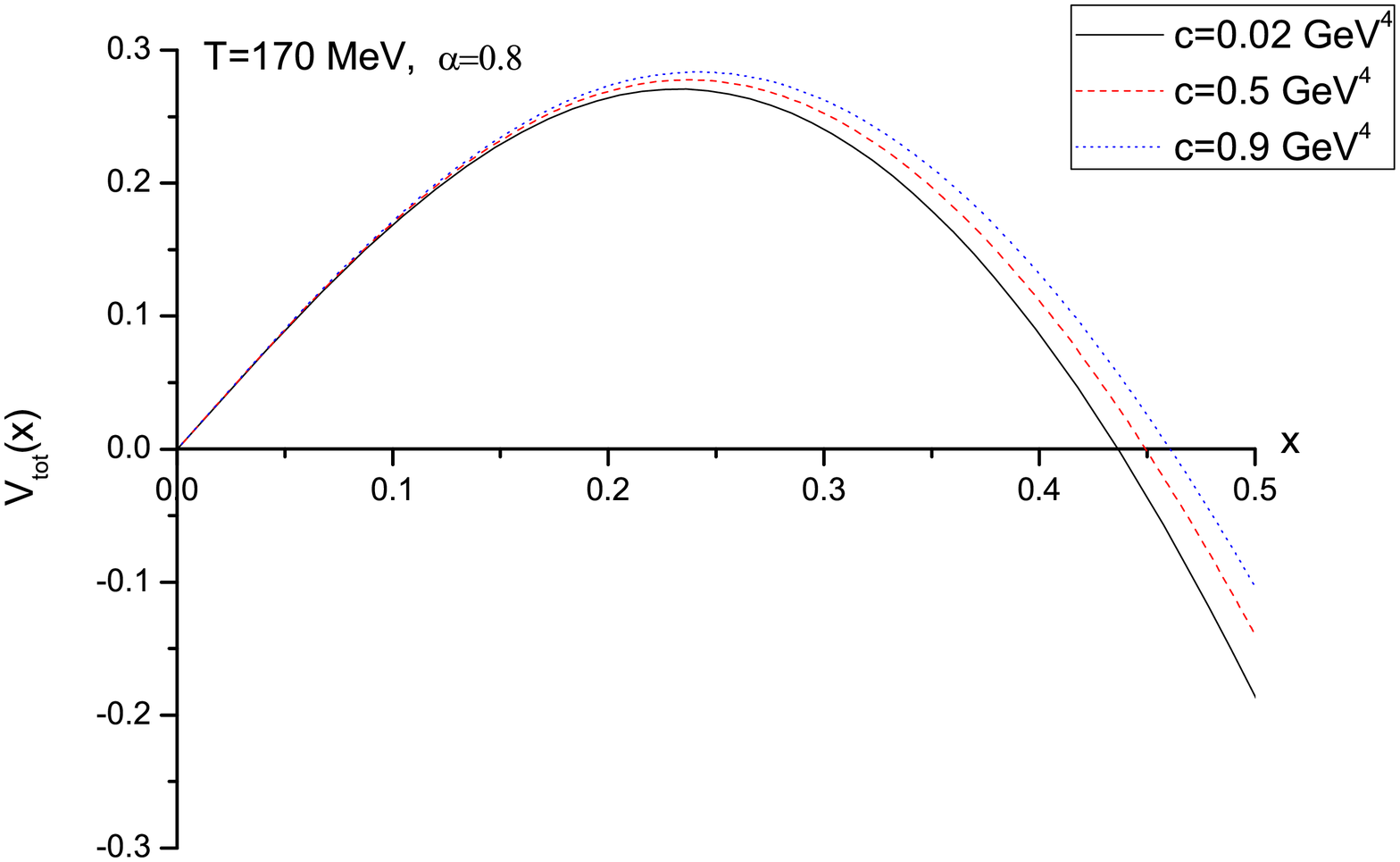}
\includegraphics[width=8cm]{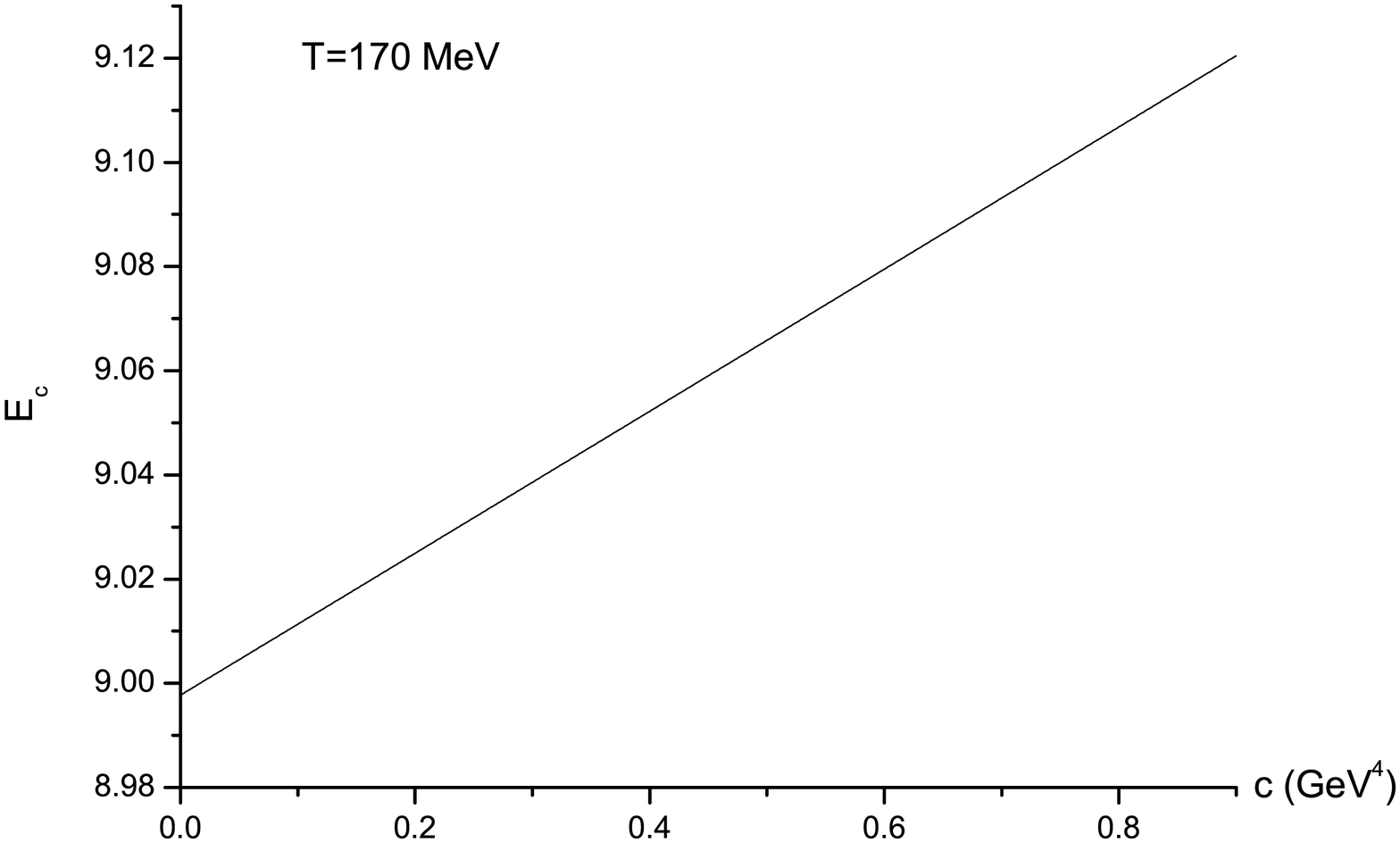}
\caption{Left: $V_{tot}(x)$ versus $x$ with fixed $\alpha=0.8$,
$T=170 MeV$ and different $c$ for the dilaton black hole
background. From top to bottom $c=0.9, 0.5, 0.02$ $GeV^4$,
respectively. Right: $E_c$ versus $c$.}
\end{figure}

\begin{figure}
\centering
\includegraphics[width=8cm]{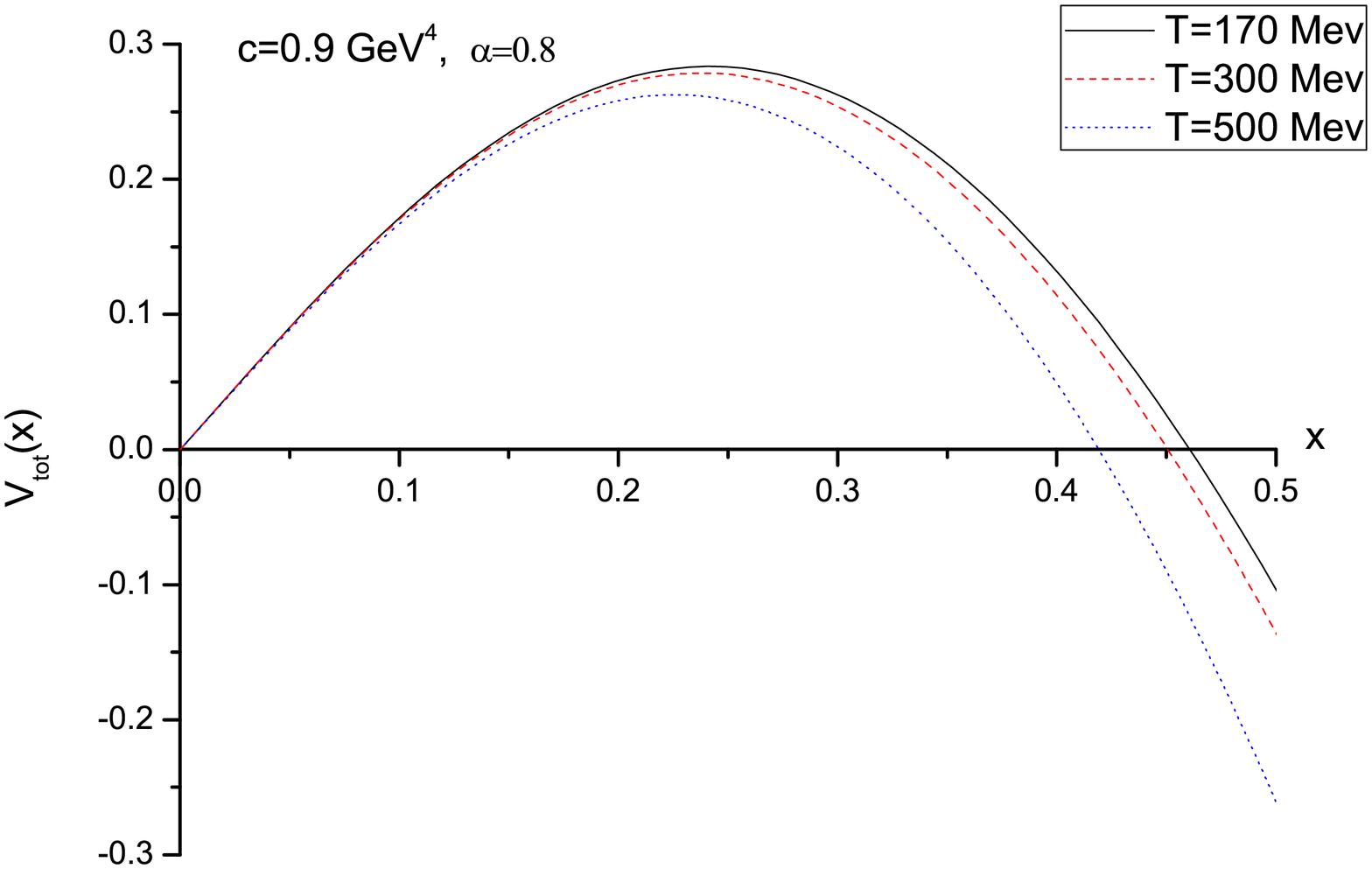}
\includegraphics[width=8cm]{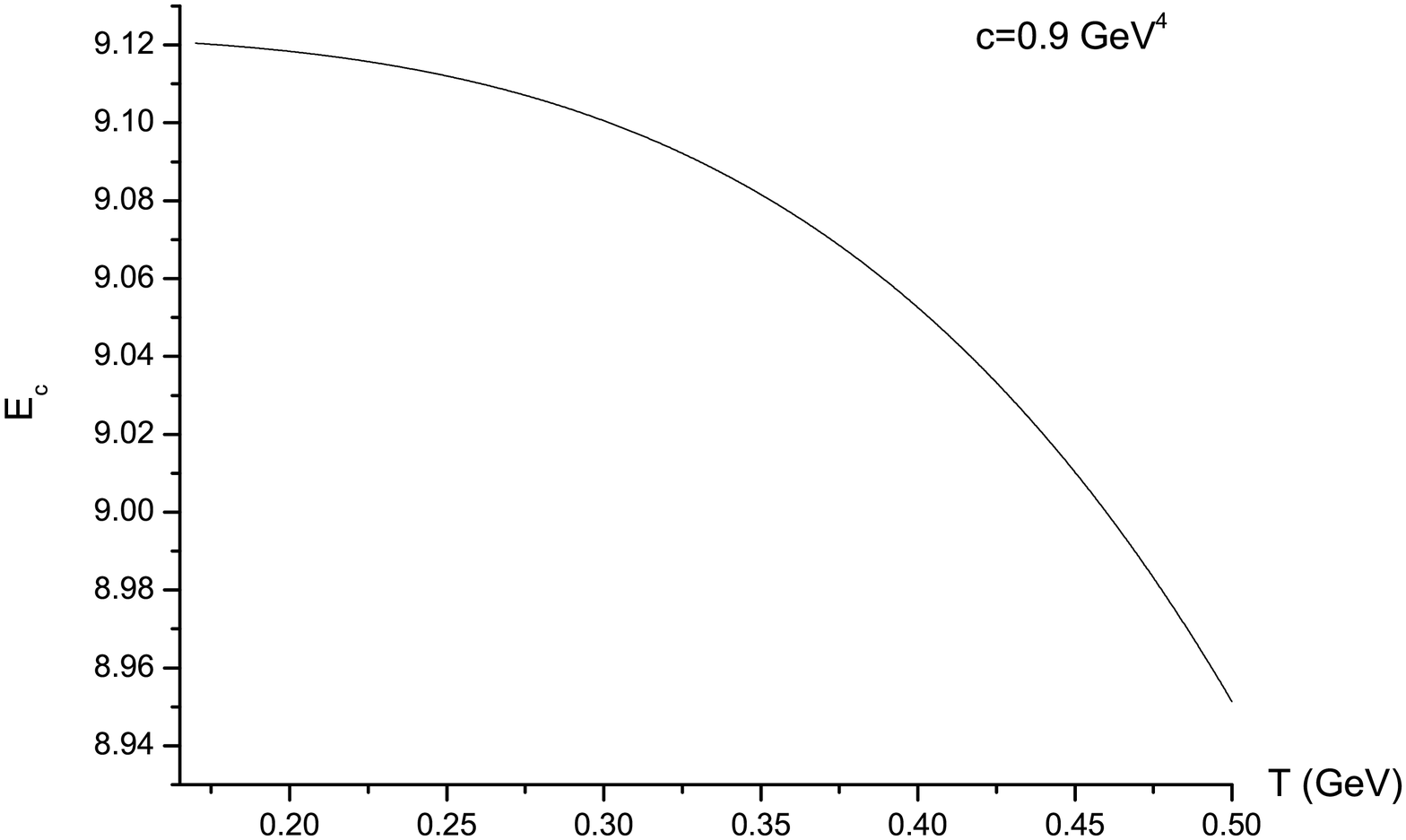}
\caption{Left: $V_{tot}(x)$ versus $x$ with fixed $\alpha=0.8$,
$c=0.9 GeV^4$ and different $c$ for the dilaton black hole
background. From top to bottom $T=170, 300, 500$ $MeV$,
respectively. Right: $E_c$ versus $T$.}
\end{figure}

\section{conclusion and discussion}
In this paper, we studied the effect of the gluon condensate on
the Schwinger effect in dilaton-wall background and dilaton black
hole background, respectively. We evaluated the electrostatic
potentials by calculating the Nambu-Goto action of a string
attaching the rectangular Wilson loop on a probe D3 brane. Also,
we determined $E_c$ from the DBI action and plotted it as a
functions of $c$ for various cases. For both backgrounds, we
observed that increasing $c$ leads to increasing the potential
barrier thus reducing the Schwinger effect. One step further, the
presence of the gluon condensate reduces the production rate, in
agreement with the finding of \cite{LS}. Also, we found the
temperature has opposite effect on the Schwinger effect.

One may wonder how gluon condensate modifies the Schwinger effect
in the investigated temperature ranges (in particular associated
with experiment)? We would like to make the following comment. It
was shown \cite{gb} that the value of $c$ drops near the
deconfinement transition. And at high temperatures, $c$ becomes
independent of $T$ and $\mu$ (the chemical potential), but when
$T$ is not very high, $c$ strongly depends on $T$ and $\mu$
\cite{zf}. Taken together, one may infer that the Schwinger effect
(or production rate) increases as $c$ decreases in the deconfined
phase, and almost won't be modified by $c$ at high temperature.
However, we could not give a concrete conclusion for intermediate
temperature or low temperature. To resolve this problem, we need
to study the competitive effects of $c$, $\mu$, $T$ (on the
Schwinger effect) as well as the relationship between the three.
We hope to report our progress in this regard in the near future.

\section{Acknowledgments}
This work is supported by the NSFC under Grant Nos. 11735007,
11705166 and the Fundamental Research Funds for the Central
Universities, China University of Geosciences (Wuhan) (No.
CUGL180402). The work of Xiangrong Zhu is supported by Zhejiang
Provincial Natural Science Foundation of China No. LY19A050001.


\end{document}